\documentclass[reprint, floatfix]{revtex4-1}

\usepackage{graphicx}

\usepackage{dcolumn}

\usepackage{amssymb}
\usepackage{amsmath}
\usepackage{epsfig}
\usepackage{chemarr}
\usepackage{url}
\usepackage{natbib}
\usepackage{color}

\usepackage[]{changes}
\definechangesauthor[name={Arvind's remarks}, color=red]{Arvind}
\definechangesauthor[name={Nachi's remarks}, color=orange]{Nachi}
\DeclareMathOperator\erf{erf}

\begin{document}

\title{Supervised learning in a mechanical system}
\author{Menachem Stern, Chukwunonso Arinze, Leron Perez, Stephanie Palmer, Arvind Murugan}

\affiliation{Department of Physics,
	University of Chicago, Chicago, IL 60637}%

\date{\today}

\begin{abstract}
Mechanical metamaterials are usually designed to show desired responses to prescribed forces. In some applications, the desired force-response relationship might be hard to specify exactly, although examples of forces and corresponding desired responses are easily available. Here we propose a framework for supervised learning in a thin creased sheet that learns the desired force-response behavior from training examples of spatial force patterns and can then respond correctly to previously unseen test forces. 
During training, we fold the sheet using different training forces and assume a learning rule that changes stiffness of creases in response to their folding strain. We find that this learning process reshapes non-linearities inherent in folding a sheet so as to show the correct response for previously unseen test forces. We study the relationship between training error, test error and sheet size which plays the role of model complexity. Our framework shows how the complex energy landscape of disordered mechanical materials can be reshaped using an iterative local learning rule.
\end{abstract}

\maketitle

The design of mechanical metamaterials usually assumes that desired force-response properties are given as a top-down specification. For example, principles of topological protection can be used to design materials where forces at specific sites lead to localized deformations~\cite{nash2015topological,bertoldi2017flexible}. Elastic networks can be pruned to exhibit allostery~\cite{rocks2017designing}, so that a deformation at one specific site is communicated to a specific distant site. In these examples and many others \cite{norman2013design,kim2019conformational,kim2019design,Pinson2017-td,stern2018shaping}, we rationally optimize design parameters, e.g., spring constants and geometry, to achieve a specified force-response relationship.

A different scenario, closely connected to supervised learning in computer science, is when the desired force-response is so complex that it cannot be specified in a top-down manner; however, it might still be easy to give examples of the desired force-response relationship. The goal is to learn or infer the right force-response relationship from these \emph{training} examples, with success evaluated on the ability to extrapolate the learned relationship to unseen \emph{test} examples.  Learning from examples in this manner offers several advantages for materials, primarily in the form of tailoring the force-response relationship to real use cases. Consider a class of spatial force patterns (e.g. exerted by cat paws), which when applied to a sheet, should fold it into one geometry and another class of force patterns (dog paws) that should result in a different folded geometry. While it might be easy to obtain and apply physical examples of forces from these classes, it is hard to mathematically list what features distinguish these two classes, especially given the large variation within the classes themselves. A learning process that automatically learns the right features from a training set can naturally solve this problem. Second, even when distinguishing features are known, learning offers a natural way of arriving at the right design parameters without need for a complex optimization algorithm. Finally, and most critically, successful learning promises a material that can show the correct response to novel inputs not seen during training.

While naturally occurring systems like neural networks~\cite{grossberg1976adaptive}, slime molds~\cite{adamatzky2010physarum}, and plant transport networks~\cite{ruder2016overview} use similar ideas to adapt their response to environmental inputs, physical supervised learning has thus far not been used to obtain functional man-made materials.
Here we propose an approach for the supervised training of a mechanical material through repeated physical actuation. We work with a model of creased thin sheets where crease stiffnesses can change as a result of repeated folding. We assume a training set, that is, a list of force patterns and desired responses. Each training example of force pattern is applied to the sheet; if the response is the desired one, as determined by a `supervisor', folded creases are allowed to soften in proportion to their folding strain. If the response is incorrect, creases stiffen instead. We then test the trained sheet by applying unseen force patterns (test examples) drawn from the same underlying distribution as the training data. We study test and training errors and thus the sheet's ability to generalize to novel patterns as a function of its size. 

Our proposal here relies on a plastic element, namely crease stiffness. Materials that stiffen or soften with strain have been demonstrated in several contexts~\cite{mullins1969softening,read1984strain,lyulin2005strain,aastrom2008strain}, including recently in the training of mechanical metamaterials~\cite{pashine2019directed}. We discuss how learning performance may be affected by limitations on the dynamic range of stiffness and other practical constraints in such materials. We hope our results here will provoke further work on how the constraints of mechanics intersect with learning.

\begin{figure*}
\includegraphics[width=1.0\linewidth]{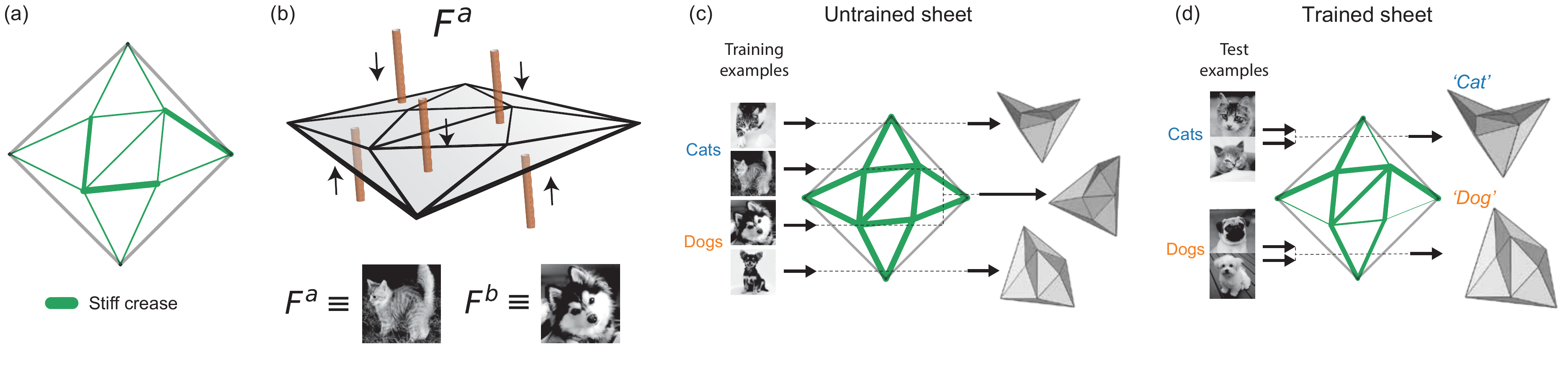}
\caption{Training thin sheets to classify spatial force patterns. a) We consider thin stiff sheets with creases whose stiffnesses (indicated by thickness of green segments) can be changed by repeated folding. b) Such sheets can fold in response to a spatial force pattern $\mathbf{F^{a}}$ applied across the sheet. To emphasize the high dimensional nature of $\mathbf{F^{a}}$, we visualize $\mathbf{F^{a}}$ as an image where the grayscale value of each pixel corresponds to the force at a particular location on the sheet. c) An untrained sheet with uniform stiffness shows random folded responses for different spatial force patterns. d) By modifying crease stiffness values, we train the sheet to classify entire classes of force patterns by showing one distinct folded response for each class.
\label{fig:Schematic}}
\end{figure*}

\section*{Results}

We demonstrate our results with a creased thin self-folding sheet (Fig.~\ref{fig:Schematic}a), which is naturally multi-stable. Our analysis can be generalized to other disordered mechanical systems, such as elastic networks~\cite{pashine2019directed}, that are also generically multi-stable.
It was previously shown that creased sheets, such as those of self-folding origami, can be folded into exponentially many discrete folded structures from the flat, unfolded state~\cite{stern2017complexity, chen2018branches}. Such exponential multi-stability can be a problem~\cite{stern2017complexity,stern2018shaping} from an engineering standpoint, as precise controlled folding is required to obtain the desired folded structure.

\begin{figure}
\includegraphics[width=1.0\linewidth]{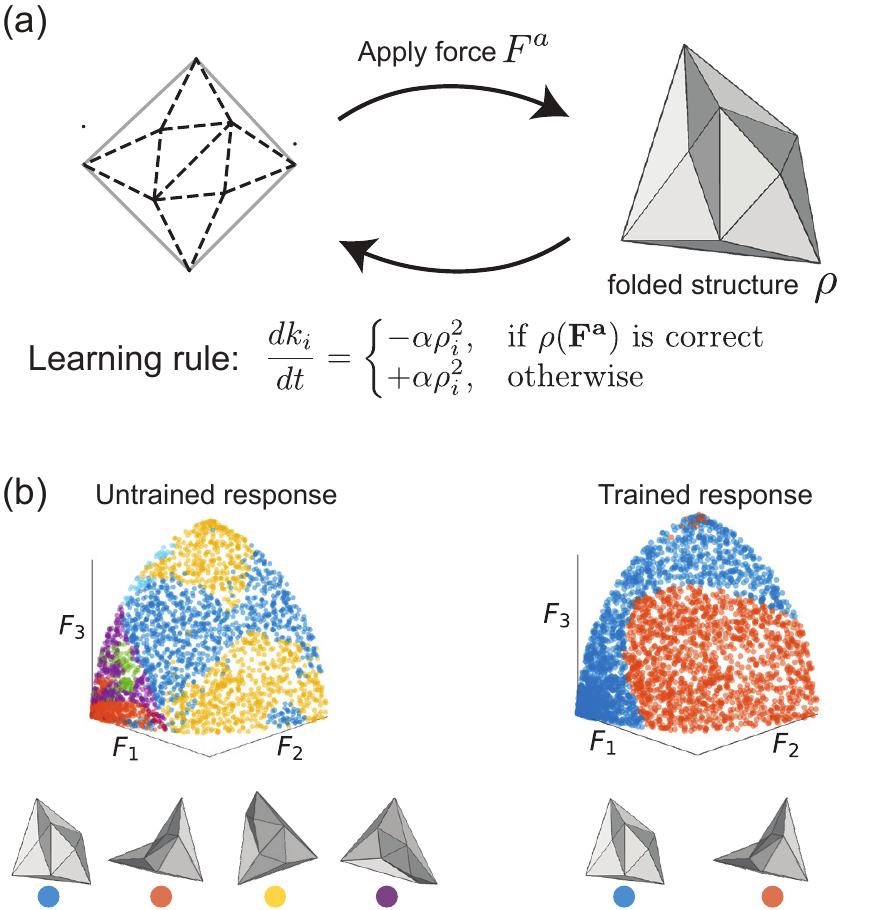}
\caption{Supervised training of thin sheets. a) A sheet with random crease geometry is folded with a training force pattern $\bm{F}^{a}$, resulting in a folded structure $\bm{\rho}$. The stiffness $k_i$ of each crease $i$ is modified according to a local learning rule; if the folded structure $\bm{\rho}$ is the desired response for $\bm{F}^{a}$ as determined by a supervisor, creases soften in proportion to their folding strain $\rho_i$. Otherwise, creases stiffen. b) This rule trains the sheet to perform the desired classification of force patterns. The untrained sheet shows multiple folded structures in response to force patterns ($2d$ cross-section of force pattern space shown). The trained sheet shows only two folded responses that mimic the desired mapping of force patterns to folded structures.
\label{fig:Rule}}
\end{figure}

Here we exploit such multi-stability to train a classifier of input force patterns. If we apply a spatial force pattern $\bm{F}$ across the flat sheet (see Fig.~\ref{fig:Schematic}a-b), the sheet will fold into a particular folded structure $\bm{\rho}(\bm{F})$, e.g., described by dihedral folding angles at each crease $\rho_i$ (see Supplementary Note 1). To emphasize the high dimensional nature of space of force patterns, in Fig.~\ref{fig:Schematic}b-d, we represent each force with a gray-scale image where each pixel can be interpreted as a force at a designated point on the sheet. The set of all force patterns $\{\bm{F}\}$ that lead to one particular folded structure $\bm{\rho}^m$ is defined as the `attractor' of folded structure $m$ in the space of force patterns (color coded regions in Fig.~\ref{fig:Rule}b). The complex attractor structure of force-response for a thin sheet naturally serves as a classifier of force patterns, albeit a random classifier (Fig.~\ref{fig:Schematic}c). The goal of the training protocol is obtain a sheet with a specific desired mapping between force patterns and folded structures (Fig.~\ref{fig:Schematic}d).

Previously, we found that the folded response to a given force pattern can be modified by changing the stiffness $k_i$ of different creases $i$ in the sheet~\cite{stern2018shaping}. Here, we employ a `supervised learning' approach to naturally tune stiffness values $k_i$ so that the sheet classifies forces as desired. Intuitively, this is done by applying examples of force patterns to the sheet and modifying crease stiffness accordingly, in a way that reinforces the correct response and discourages incorrect folding (Fig.~\ref{fig:Rule}a). Such training, carried out iteratively for different force pattern examples, has the effect of morphing the attractor structure to better approximate the desired response (Fig.\ref{fig:Rule}b).

Consider two distributions of force patterns, each designated as a particular class (e.g. `cats' and `dogs'). An example is shown in Fig.~\ref{fig:Caps}a (top) where the two classes of forces are defined by spherical caps in force space. We define all forces belonging to these distributions by $S^{\text{dog}} = \{ \bm{F} |  \bm{F} \cdot \bm{F}_{\text{dog}} \geq D, \bm{F} \cdot \bm{F}_{\text{dog}} > \bm{F} \cdot \bm{F}_{\text{cat}} \}$, and similarly for $S^{\text{cat}}$. Here $D=0.6$ sets the size of the caps. (In Fig.~\ref{fig:Caps}a, $S^{\text{dog}}$ is blue and $S^{\text{cat}}$ is orange.)

Assume we are given two sets of labeled force patterns as training examples $\mathcal{F}^{\text{dog}} = \{\bm{F}\in S^{\text{dog}}\}$ , $\mathcal{F}^{\text{cat}} = \{\bm{F}\in S^{\text{cat}}\}$, each with $n$ training force patterns (in Fig.~\ref{fig:Caps}a (bottom) we sample sets with $n=20$). Together, $\mathcal{F}^{\text{dog}}$ and $\mathcal{F}^{\text{cat}}$ are defined as the \textit{training set}. We desire all forces in $S^{\text{dog}}$ to result in one common folded structure, while all forces in $S^{\text{cat}}$ fold the sheet to a distinct but common folded structure. While $S^{\text{dog}}$, $S^{\text{cat}}$ are separable in some $2d$ projection of force space, learning is non-trivial since the sheet must learn the $2$ dimensions in which these distributions are separable.

\subsection*{A mechanical supervised training protocol}

In our training protocol, each of the training force patterns $\bm{F}^a$ is applied to the sheet in sequence, to obtain a folded structure $\bm{\rho}(\bm{F}^a)$. A supervisor determines whether the resulting folded state $\bm{\rho}(\bm{F}^a)$ is correct or incorrect by comparing it to a reference state $\bm{\rho}_{\text{ref}}(\bm{F})$ for those classes. (The reference state can be selected in several ways. Here, we average the response of the untrained sheet on training examples in each class; see Supplementary Note 2.)

We then apply the following learning rule that stiffens or softens each crease in proportion to folding in that crease,
\begin{equation}
  \frac{dk_i}{dt} = \left \{
  \begin{aligned}
    &- \alpha \rho_i^r, && \text{if } \bm{\rho}(\bm{F}^a) \text{ is correct} \\
    &+ \alpha \rho_i^r, && \text{otherwise}
  \end{aligned} \right.
  \label{eq:LearningRule}
\end{equation}
for the stiffness $k_i$ of each crease $i$. $\alpha$ is a \textit{learning rate}, setting how fast stiffness values $k_i$ are updated due to training examples. $r$ models non-linearities in strain-based softening or stiffening of materials; we use $r=2$. Such plasticity is experimentally seen in several materials~\cite{deng1994effect,hong2005dynamic,gui2001aging}; we discuss other learning rules and experimental constraints later.

After each round of training the pattern is unfolded back to the flat state. The same supervised learning step is then repeated in sequence for all training force patterns. A training epoch is defined as one pass through the entire training set.

\begin{figure}
\includegraphics[width=1.0\linewidth]{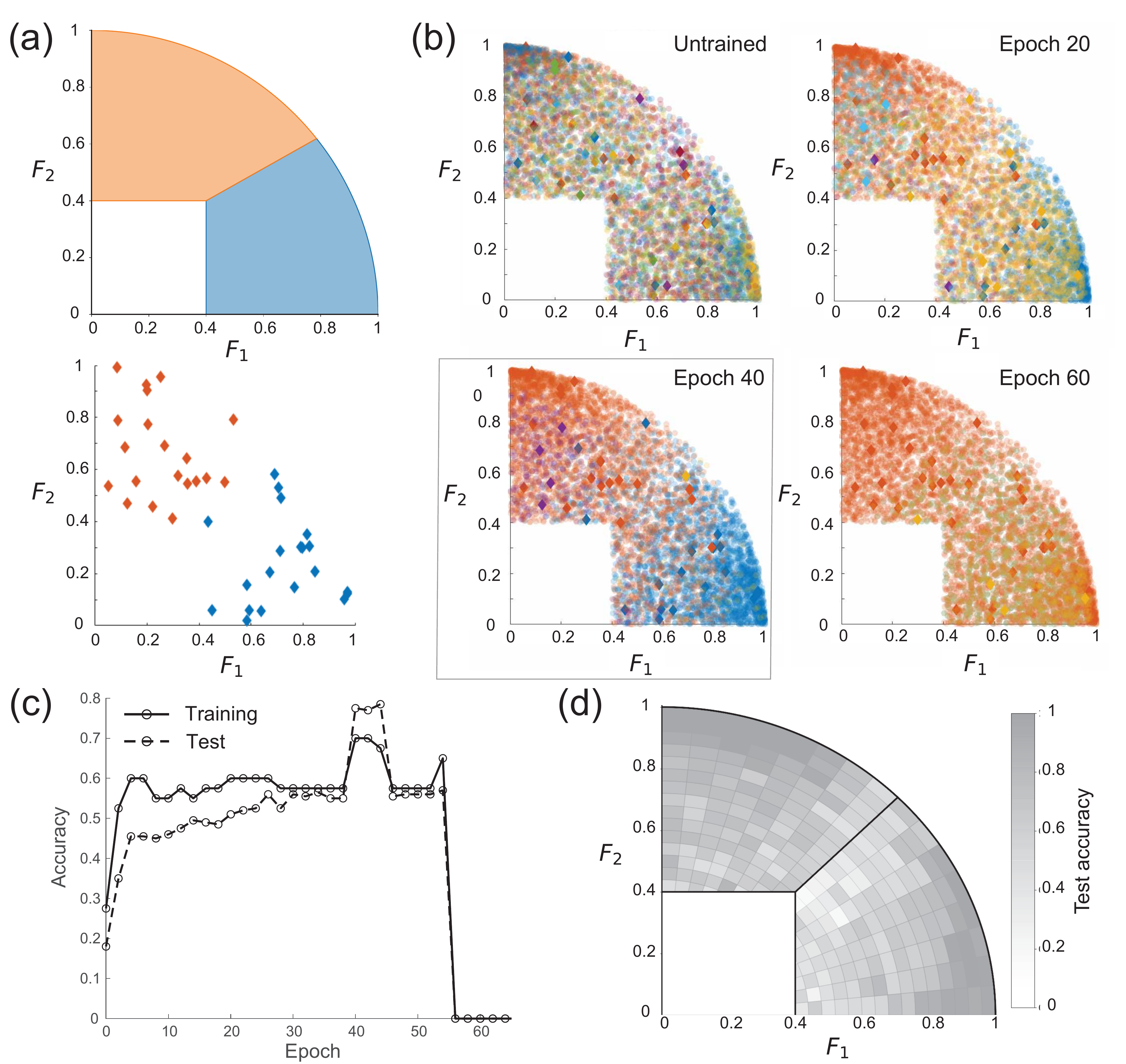}
\caption{Supervised learning of cap-like force distributions. a) We define distributions $S^{\text{dog
}}$ (blue) and $S^{\text{dog
}}$ (orange) of force patterns as two spherical caps in the space of applied forces (2d projection shown). Training examples (diamonds) are drawn from both  distributions. b) An untrained sheet folds into many distinct folded structures (different colors) in response to applied force patterns. As training progresses, most force patterns are classified as either blue or orange according to the cap they belong to. When over-trained, all force patterns result in only one folded structure (orange). c) The trained sheet reaches peak performance after $\sim 40$ epochs of supervised training (i.e., passes through the training examples). The trained sheet not only classifies the training set correctly (training accuracy), but generalizes to unseen test force patterns (test accuracy). d) The trained sheet is highly accurate when classifying force patterns near the center of the spherical caps, but less accurate close to the true decision boundary between the distributions.
\label{fig:Caps}}
\end{figure}

We find that as training proceeds, the number of observed folded structures decreases (fewer colors), and nearly all training force patterns fold the sheet into the `blue' or the `orange' labeled structures after epoch $40$ (diamonds in Fig.~\ref{fig:Caps}b). The fraction of training force patterns that fold the sheet into the correct structure is defined as the \textit{training accuracy}.

However, a successfully trained sheet should correctly classify previously unseen `test` force patterns, sampled from the same distributions. We tested the trained sheet by applying such test examples drawn from the caps $S^{\text{dog}},S^{\text{cat}}$ and recording the resulting folded structure. In analogy to the training sets, the fraction of test examples yielding the correct folded structure is defined as the \textit{test accuracy}. High test accuracy is observed (Fig.~\ref{fig:Caps}c,d) ($\sim 80\%$ of the test examples classified correctly); thus the sheet generalizes and is able to have the right response to novel test force patterns through the changes induced by training examples.

\begin{figure}
\includegraphics[width=1.0\linewidth]{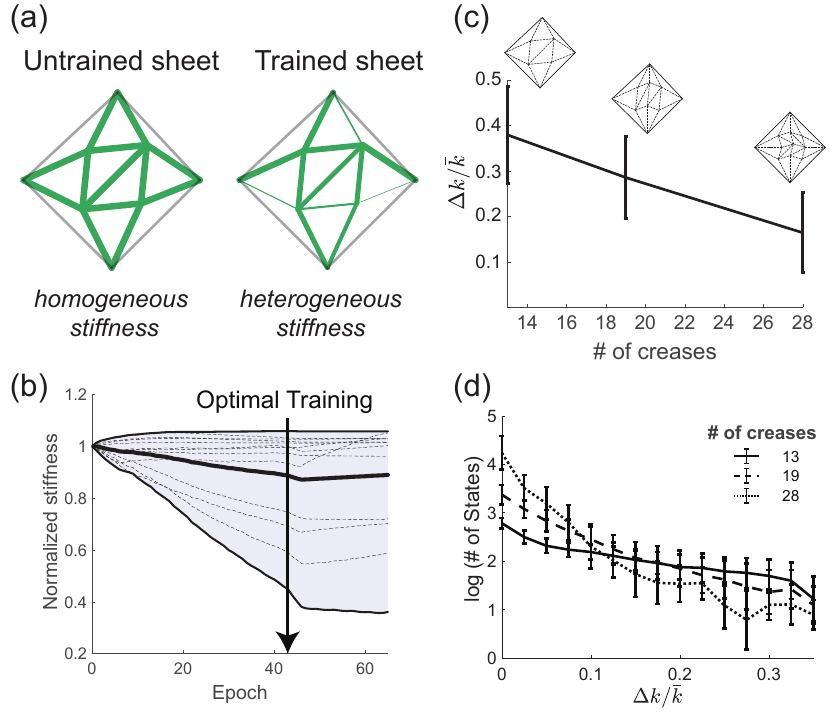}
\caption{Training increases the variance of crease stiffness across the sheet. a) Untrained sheets have a homogeneous distribution of crease stiffnesses, while trained sheets have heterogeneous stiffness profiles (width of green lines). b) As the sheet is trained, the stiffness of different creases changes to different extents, such that the variance in stiffness values grows over training time (envelope shows the least and most stiff creases). c) Larger sheets with more creases require smaller variance in their stiffness values for optimal training. d) An untrained sheet starts with exponentially many available folded structures. During training, the number of available folded structures decreases exponentially with increasing stiffness variance $\Delta k^2$, allowing the sheet to classify a few distinct classes.
\label{fig:KVariance}}
\end{figure}

\subsection*{Heterogeneous crease stiffness}

Our learning rule facilitates classification by creating heterogeneous crease stiffness across the sheet (Fig.~\ref{fig:KVariance}a). Indeed, as training proceeds, we find that the variance $\Delta k^2$ of stiffness grows (Fig.~\ref{fig:KVariance}b). If the sheet is trained beyond the optimal point, the stiffness variance still grows, but the classifier eventually fails, as seen in Fig.~\ref{fig:Caps}b-c. The failure mode of over-training is typically that all forces fold the sheet into a single folded structure, resulting in no classification.

We can understand this relationship between heterogeneous stiffness ($\Delta k$) and training using a simple model. A heterogeneous crease stiffness profile $\bm{k}$ with high stiffness $k_i$ in crease $i$ but no stiffness elsewhere, will lift the energy of structures $\bm{\rho}$ with small folding $\rho_i$ in crease $i$ less than structures with large $\rho_i$. Hence heterogeneous $\bm{k}$ can raise the energy of select structures, reducing their attractor size, while other structures remain low in energy and grow in attractor size. If we assume that folding angles $\bm{\rho}^a$ of structure $a$ are randomly distributed (verified earlier in~\cite{Pinson2017-td}) and assume a random stiffness pattern with standard deviation $\Delta k$, the energies $\sum_i k_i \rho^2_i$ of different structures will be distributed as a Gaussian with mean $\mu=\alpha \bar{k}$ and standard deviation $\sigma = \beta \Delta k$ where $\bar{k}$ is the mean stiffness, and $\alpha ,\beta$ some numerical parameters.

If structures above energy $E_F$ are inaccessible to folding, the number of accessible folded structures is,
\begin{align}
\#(\Delta k,N)  \sim 2^N [1-\erf(\frac{\alpha \bar{k} - E_F}{\beta \Delta k})] 
\label{eq:NumberAttractors}.
\end{align}

Hence the number of surviving folded structures should decrease fast with $\Delta k$. This effect is indeed observed for trained origami sheets of different sizes (Fig.~\ref{fig:KVariance}d). From numerical exploration of the energy distributions in this model, we find that $\alpha$ is a constant regardless of sheet size, while $\beta\sim N^{-0.5}$ shrinks with sheet size (central limit theorem). Using this form of $\beta$ in Eq.~\ref{eq:NumberAttractors} predicts that the elimination of structures happens at a lower $\Delta k$ for larger sheets, consistent with our results in Fig.~\ref{fig:KVariance}c-d.

We conclude that as the training protocol proceeds, the stiffness variance $\Delta k^2$ grows, and the number of available folded structures decreases. The last surviving folded structures, reinforced by the learning rule of Eq~(\ref{eq:LearningRule}), classify the force distributions correctly. Thus, the learning process merges attractors of the untrained sheet such that the surviving attractors recapitulate features of the desired force-fold mapping.

\subsection*{Generalization and sheet size}

\begin{figure}
\includegraphics[width=1.0\linewidth]{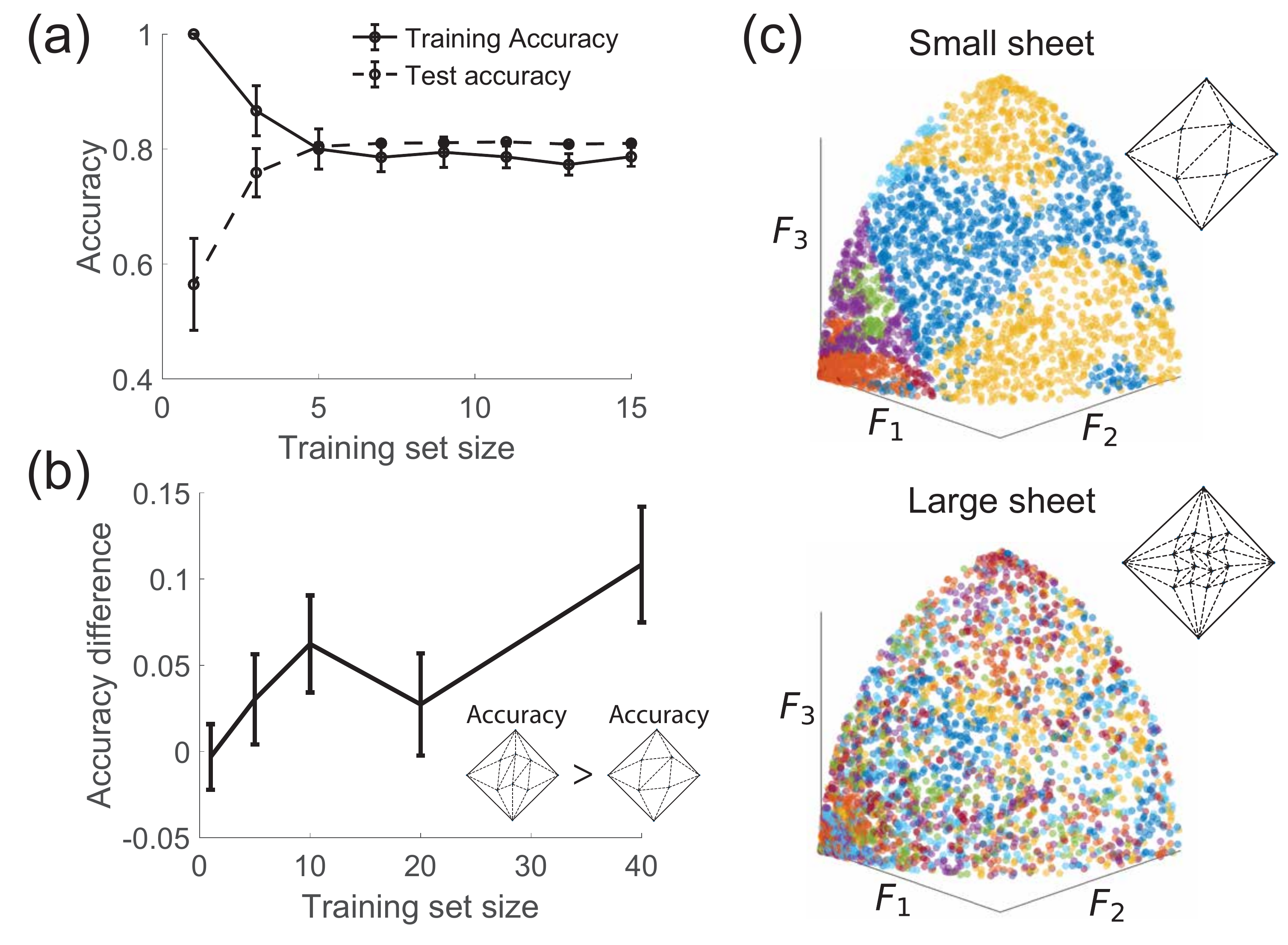}
\caption{Effect of training set size and sheet size on test accuracy. a) With fewer distinct training examples, training accuracy is high but test accuracy is low (overfitting). Increasing the number of training examples improves test accuracy, at the expense of training accuracy. b) Sheets with more creases show larger improvements in test accuracy with increasing number training examples, as expected of complex models with more fitting parameters. c) A small untrained sheet (13 creases) shows $\sim 10$ folded structures (color coded) in response to different force patterns. A larger sheet (49 creases) sheet shows $\sim 400$ folded structures instead, each with smaller attractor regions in the space of force patterns. Consequently, larger sheets can create more flexible classification surfaces by combining smaller attractor regions; such complex models with more fitting parameters require more training examples to avoid overfitting.
\label{fig:Size}}
\end{figure}

\begin{figure*}
\includegraphics[width=1.0\linewidth]{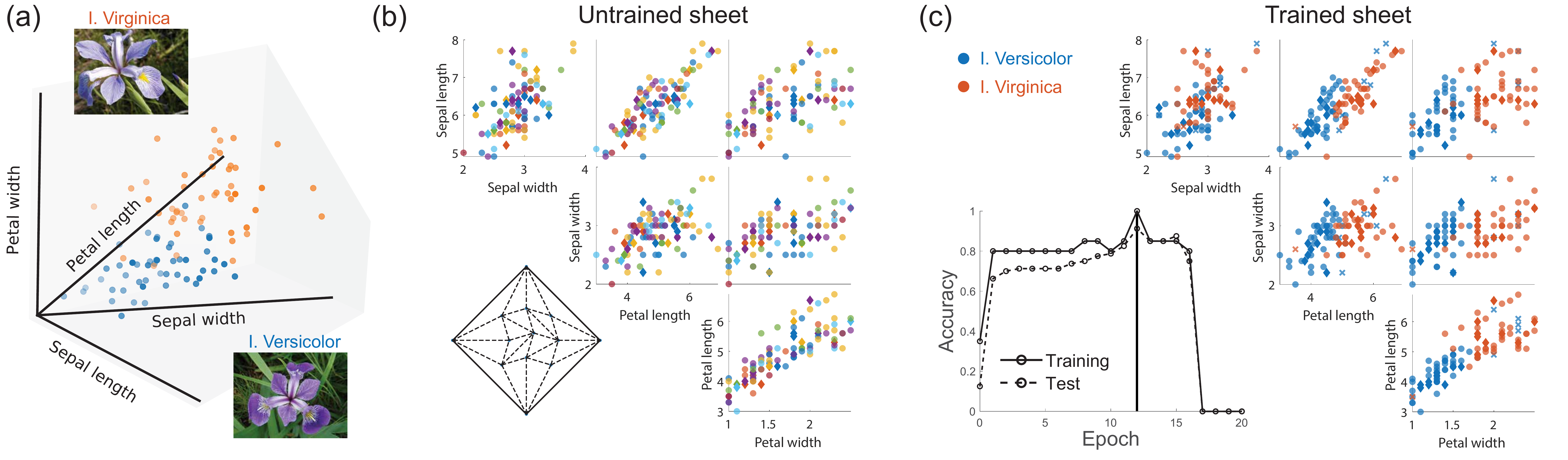}
\caption{Training sheets to classify Iris specimens. a) We train a sheet to classify individual Iris specimens as one of two species based on petal and sepal lengths and widths~\cite{Iris}. We translate these four measurements into a spatial pattern of forces applied to the sheet. b) Folding response of an untrained (28 crease) sheet due to force patterns derived from the Iris data (shown in every cross section). c) The sheet is trained using $10$ random examples (diamonds) of each species from the database \cite{Iris} and then tested on $80$ unseen test examples (circles). Matrix shows the classification of Iris flowers at optimal training of the sheet ($91\%$ test accuracy; mistakes denoted by x).
\label{fig:Iris}}
\end{figure*}

Statistical learning theory~\cite{Vapnik2013-jn} suggests that two critical parameters set the quality of learning: (1) the number of training examples seen, (2) complexity of the learning model. An increased number of training examples usually decreases training accuracy. However, test accuracy - i.e., the response to novel examples or the ability to generalize - improves. Furthermore, the improvement of test accuracy is larger for complex models with more fitting parameters. Intuitively, complex models `overfit' details of small training sets, and thus show low test accuracy even if training accuracy is high. 

Our sheets exhibit signatures of these learning theory results, with the size of the sheet (number of creases) playing the role of model complexity. For a sheet of fixed size, trained on the distributions of Fig.~\ref{fig:Caps}, we observe that increasing the number of training examples increases test accuracy and decreases training accuracy (Fig.~\ref{fig:Size}a). In Fig.~\ref{fig:Size}b, we find the test accuracy of larger sheets with more creases improves more dramatically with the size of the training set, compared to smaller sheets.

These results suggest that sheets with more creases correspond to more complex classification models (e.g., a neural network with more neurons). For example, crease stiffnesses are the learnable parameters in our approach; hence increasing their number amounts to using a training model with more parameters. Further, untrained sheets with more creases support exponentially more folded structures~\cite{stern2017complexity, chen2018branches} as shown by the color coded force-to-folded-structure relationship in Fig.~\ref{fig:Size}c. 
The training protocol achieves correct classification by merging different colored regions. Thus, larger sheets can approximate more complex decision boundaries by combining the smaller regions shown in Fig.~\ref{fig:Size}c., and thus act as more complex models to be favored when the amount of training data is large. In the Discussions, we use these results to contrast memory and learning in mechanical systems.

\subsection*{Complex classification problems}

The attractor structure of disordered thin sheets is complex, and contains an exponential number of attractors. Hence, sheets can be expected to learn more complex features than those shown previously in Fig.~\ref{fig:Caps}.

We tested our learning protocol on the classic Iris data set~\cite{Iris2}, used extensively in the past to benchmark classification algorithms. This data set reports four measurements - length and width of petals and sepals - for individual specimens of different Iris species. While different Iris species cannot be distinguished by any one of these properties, we wanted to test if our sheet can learn the combination of features needed to distinguish species.

We picked the two most similar species in this data set, \emph{Iris Versicolor} and \emph{Iris Virginica} (Fig.~\ref{fig:Iris}a). We translated the four flower measurements to four force components applied to a sheet (see Supplementary Note 4). We then applied our training algorithm with a training set consisting of $10$ examples of \emph{I. Versicolor} and \emph{I. Virginica} (diamonds in Fig.~\ref{fig:Iris}c). The resulting trained sheet was tested on $80$ \emph{unseen} examples of these species; the trained sheet was able to identify the species of $91\%$ of previously unseen specimens correctly.

We have tested our training protocol on more complex, higher dimensional distributions (Supplementary Note 3). For example, we used the folding behavior of one thin sheet (the master) as the target behavior for another thin sheet with a distinct crease geometry. We find that the trained sheet is able to correctly predict the response of the master sheet to forces not seen during training. Thus, using our training protocol, sheets can learn and generalize complex force-to-folded-response maps from examples.

\subsection*{Experimental considerations}

Our learning framework requires materials that can plastically stiffen or soften when strained repeatedly~\cite{kuder2013variable}. Several such materials and structures are known, including shape memory polymers~\cite{mcknight2005variable,mather2009shape}, shape memory alloys (Nitinol)~\cite{cross1969nitinol}, and fluidic flexible matrix composites~\cite{philen2007fluidic}. Such materials have the advantage of truly variable, user controlled stiffness. Other materials can show a plastic change in stiffness in response to aging under strain, such as Ethylene Vinyl Acetate (EVA) foam~\cite{jin2010uv} and thermoplastic Polyurethane~\cite{boubakri2010impact}. EVA was used recently~\cite{pashine2019directed} to show such behavior in a mechanical system trained for auxetic response.

The specific learning rule used in the paper requires the ability to soften or stiffen depending on the supervisor's judgement of outcome. Such a learning rule can be implemented by materials that stiffen under strain in one condition (say, high temperature, low pH) but soften under strain in another condition.

\begin{figure}
\includegraphics[width=1.0\linewidth]{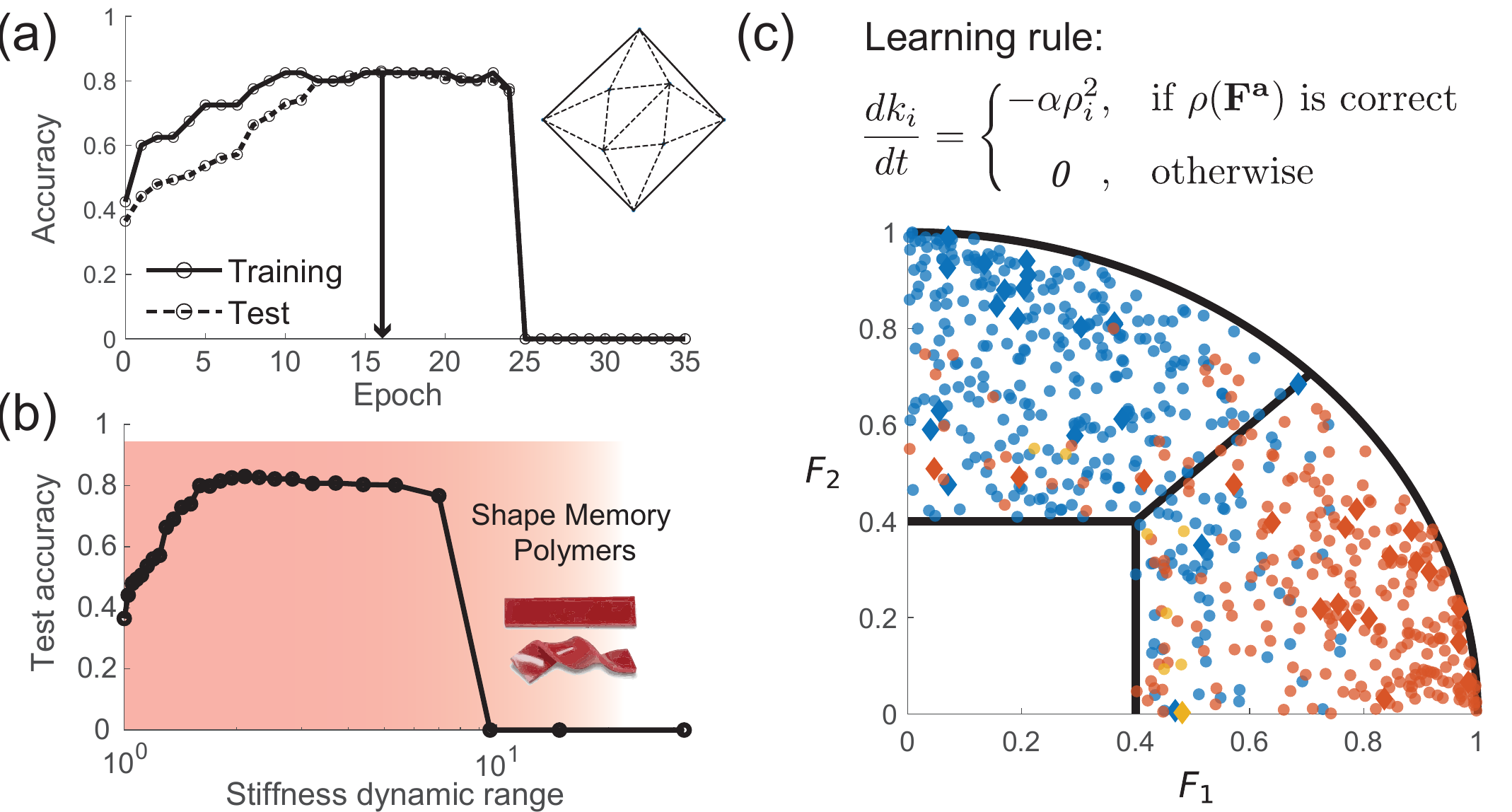}
\caption{Learning is successful even with simplified training rules and experimentally realizable stiffness range. a) A sheet ($13$ crease) trained on the classification problem of Fig.~\ref{fig:Caps}, with a simplified, experimentally viable learning rule shown in (c). b) At peak training, the dynamic range of crease stiffness values is $\sim 2$, well within the ranges supported by existing shape memory polymers (red filled region)~\cite{mcknight2005variable}.  c) Trained sheet reaches a peak accuracy of $\sim 80\%$ on test force examples (circles).
\label{fig:OneSidedRule}}
\end{figure}

However, the results here also hold for simpler learning rules, e.g., that only require plastic softening under strain. For example, we can modify the learning rule (Eq.~\ref{eq:LearningRule}) to :
\begin{equation}
  \frac{dk_i}{dt} = \left \{
  \begin{aligned}
    &- \alpha \rho_i^2, && \text{if } \bm{\rho}(\bm{F}) \text{ is correct} \\
    &\ \ \ \ 0, && \text{otherwise}
  \end{aligned} \right.
\end{equation}

Such a rule is easily implemented with a strain-softening material with no stiffening needed. For example, if the folded outcome $\bm{\rho}(\bm{F})$ is judged correct, we hold the sheet in the folded state $\bm{\rho}(\bm{F})$ for a  longer length of time (allowing significant softening) than when the outcome is judged incorrect (no softening). We tested this simplified learning rule for the classification problem in  Fig.~\ref{fig:Caps}; we find similar accuracy as earlier (see Fig.~\ref{fig:OneSidedRule}a).

Another significant experimental constraint is the dynamic range of crease stiffnesses $k_i$ achievable in real materials without failure or fracture at the creases. Fortunately, we find that for well trained sheets, the difference in crease stiffness is  moderate (Fig.~\ref{fig:KVariance}c), and does not exceed $30\%$ of the mean stiffness value for a medium sized ($28$ crease) sheet. Fig.~\ref{fig:OneSidedRule}b shows that our required dynamic range in stiffness is within the range for experimentally available materials such as shape memory polymers~\cite{mcknight2005variable,lagoudas2008shape}.

Finally, another failure mode for our training protocol is overtraining. While the variance in crease stiffness $\Delta k^2$ is critical to eliminate attractors, overtraining can result in a sheet with only one folded structure. Our analysis, presented in Fig.~\ref{fig:KVariance}, suggests that large sheets should be easier to train experimentally since the stiffness variance needed is more moderate, while the transition to overtraining does not become much more rapid.

\section*{Discussion}

In this work, we have demonstrated the supervised training of a mechanical system, a thin creased sheet, to classify input force patterns. As required for learning, the trained sheet not only shows the correct response for training forces, but can generalize and show the correct response to unseen test examples of forces. We studied the relationship between training error, test error, and the size of the sheet which plays the role of model complexity in supervised learning~\cite{Vapnik2013-jn}.

We can contrast the learning framework here with that of memory formation in mechanical systems~\cite{keim2019memory}. A robust memory implies a trained model that shows the correct response for all of the training examples (i.e., low training error), even at the expense of overfitting such training data. In contrast, in learning, we seek a model that generalizes and responds correctly to unseen examples (i.e., low test error), even at some expense of training error. With this view, large sheets trained with a small number of training examples can serve as memory while smaller sheets with more training examples lead to generalization, and hence learning.

Supervised training of mechanical systems offers advantages over traditional mechanical design. On a practical level, a material with an intrinsic mechanical learning rule can be trained by an end-user rather than an expert designer, according to the task at hand. In the case of sheets presented here, the same sheet can be used to classify different data distributions depending on how it was trained. Such properties are highly sought after, e.g. in adaptive robotics~\cite{paik2012robotic}.  Finally, as learning allows generalization, materials can be trained to show a desired force-response behavior even if some examples of use are not available at the time of training.

\section*{Acknowledgments}
We thank Daniel Hexner, Nathan Kutz, Sidney Nagel, Paul Rothemund and Thomas Witten for insightful discussions. We acknowledge NSF-MRSEC 1420709 for funding and the University of Chicago Research Computing Center for computing resources.


\begin{thebibliography}{10}
	
	\bibitem{nash2015topological}
	Lisa~M Nash, Dustin Kleckner, Alismari Read, Vincenzo Vitelli, Ari~M Turner,
	and William~TM Irvine.
	\newblock Topological mechanics of gyroscopic metamaterials.
	\newblock {\em Proceedings of the National Academy of Sciences},
	112(47):14495--14500, 2015.
	
	\bibitem{bertoldi2017flexible}
	Katia Bertoldi, Vincenzo Vitelli, Johan Christensen, and Martin van Hecke.
	\newblock Flexible mechanical metamaterials.
	\newblock {\em Nature Reviews Materials}, 2(11):17066, 2017.
	
	\bibitem{rocks2017designing}
	Jason~W Rocks, Nidhi Pashine, Irmgard Bischofberger, Carl~P Goodrich, Andrea~J
	Liu, and Sidney~R Nagel.
	\newblock Designing allostery-inspired response in mechanical networks.
	\newblock {\em Proceedings of the National Academy of Sciences},
	114(10):2520--2525, 2017.
	
	\bibitem{norman2013design}
	Don Norman.
	\newblock {\em The design of everyday things: Revised and expanded edition}.
	\newblock Basic books, 2013.
	
	\bibitem{kim2019conformational}
	Jason~Z Kim, Zhixin Lu, Steven~H Strogatz, and Danielle~S Bassett.
	\newblock Conformational control of mechanical networks.
	\newblock {\em Nature Physics}, page~1, 2019.
	
	\bibitem{kim2019design}
	Jason~Z Kim, Zhixin Lu, and Danielle~S Bassett.
	\newblock Design of large sequential conformational change in mechanical
	networks.
	\newblock {\em arXiv preprint arXiv:1906.08400}, 2019.
	
	\bibitem{Pinson2017-td}
	Matthew~B Pinson, Menachem Stern, Alexandra Carruthers~Ferrero, Thomas~A
	Witten, Elizabeth Chen, and Arvind Murugan.
	\newblock Self-folding origami at any energy scale.
	\newblock {\em Nat. Commun.}, 8:15477, 18~May 2017.
	
	\bibitem{stern2018shaping}
	Menachem Stern, Viraaj Jayaram, and Arvind Murugan.
	\newblock Shaping the topology of folding pathways in mechanical systems.
	\newblock {\em Nat. Commun.}, 9(1):4303, 2018.
	
	\bibitem{grossberg1976adaptive}
	Stephen Grossberg.
	\newblock Adaptive pattern classification and universal recoding: Ii. feedback,
	expectation, olfaction, illusions.
	\newblock {\em Biological cybernetics}, 23(4):187--202, 1976.
	
	\bibitem{adamatzky2010physarum}
	Andrew Adamatzky.
	\newblock {\em Physarum machines: computers from slime mould}, volume~74.
	\newblock World Scientific, 2010.
	
	\bibitem{ruder2016overview}
	Sebastian Ruder.
	\newblock An overview of gradient descent optimization algorithms.
	\newblock {\em arXiv preprint arXiv:1609.04747}, 2016.
	
	\bibitem{mullins1969softening}
	Leonard Mullins.
	\newblock Softening of rubber by deformation.
	\newblock {\em Rubber chemistry and technology}, 42(1):339--362, 1969.
	
	\bibitem{read1984strain}
	H\_E Read and GA~Hegemier.
	\newblock Strain softening of rock, soil and concrete—a review article.
	\newblock {\em Mechanics of Materials}, 3(4):271--294, 1984.
	
	\bibitem{lyulin2005strain}
	AV~Lyulin, B~Vorselaars, MA~Mazo, NK~Balabaev, and MAJ Michels.
	\newblock Strain softening and hardening of amorphous polymers: Atomistic
	simulation of bulk mechanics and local dynamics.
	\newblock {\em EPL (Europhysics Letters)}, 71(4):618, 2005.
	
	\bibitem{aastrom2008strain}
	Jan~A {\AA}str{\"o}m, PB~Sunil Kumar, Ilpo Vattulainen, and Mikko Karttunen.
	\newblock Strain hardening, avalanches, and strain softening in dense
	cross-linked actin networks.
	\newblock {\em Physical Review E}, 77(5):051913, 2008.
	
	\bibitem{pashine2019directed}
	Nidhi Pashine, Daniel Hexner, Andrea~J Liu, and Sidney~R Nagel.
	\newblock Directed aging, memory and nature's greed.
	\newblock {\em arXiv preprint arXiv:1903.05776}, 2019.
	
	\bibitem{stern2017complexity}
	Menachem Stern, Matthew~B Pinson, and Arvind Murugan.
	\newblock The complexity of folding self-folding origami.
	\newblock {\em Phys. Rev. X}, 7(4):041070, 2017.
	
	\bibitem{chen2018branches}
	Bryan Gin-ge Chen and Christian~D Santangelo.
	\newblock Branches of triangulated origami near the unfolded state.
	\newblock {\em Physical Review X}, 8(1):011034, 2018.
	
	\bibitem{deng1994effect}
	YW~Deng, TL~Yu, and CH~Ho.
	\newblock Effect of aging under strain on the physical properties of
	polyester--urethane elastomer.
	\newblock {\em Polymer journal}, 26(12):1368, 1994.
	
	\bibitem{hong2005dynamic}
	Seong-Gu Hong, Keum-Oh Lee, and Soon-Bok Lee.
	\newblock Dynamic strain aging effect on the fatigue resistance of type 316l
	stainless steel.
	\newblock {\em International Journal of Fatigue}, 27(10-12):1420--1424, 2005.
	
	\bibitem{gui2001aging}
	Shan~Zi Gui and Yukuo Nanzai.
	\newblock Aging in quenched poly (methyl methacrylate) under inelastic tensile
	strain.
	\newblock {\em Polymer journal}, 33(5):444, 2001.
	
	\bibitem{Vapnik2013-jn}
	Vladimir Vapnik.
	\newblock {\em The nature of statistical learning theory}.
	\newblock Springer science \& business media, 2013.
	
	\bibitem{Iris}
	Ronald~A Fisher.
	\newblock The use of multiple measurements in taxonomic problems.
	\newblock {\em Annals of eugenics}, 7(2):179--188, 1936.
	
	\bibitem{Iris2}
	Anil~K Jain, Robert P.~W. Duin, and Jianchang Mao.
	\newblock Statistical pattern recognition: A review.
	\newblock {\em IEEE Transactions on pattern analysis and machine intelligence},
	22(1):4--37, 2000.
	
	\bibitem{kuder2013variable}
	Izabela~K Kuder, Andres~F Arrieta, Wolfram~E Raither, and Paolo Ermanni.
	\newblock Variable stiffness material and structural concepts for morphing
	applications.
	\newblock {\em Progress in Aerospace Sciences}, 63:33--55, 2013.
	
	\bibitem{mcknight2005variable}
	Geoff McKnight and Chris Henry.
	\newblock Variable stiffness materials for reconfigurable surface applications.
	\newblock In {\em Smart Structures and Materials 2005: Active Materials:
		Behavior and Mechanics}, volume 5761, pages 119--126. International Society
	for Optics and Photonics, 2005.
	
	\bibitem{mather2009shape}
	Patrick~T Mather, Xiaofan Luo, and Ingrid~A Rousseau.
	\newblock Shape memory polymer research.
	\newblock {\em Annual Review of Materials Research}, 39:445--471, 2009.
	
	\bibitem{cross1969nitinol}
	William~B Cross, Anthony~H Kariotis, and Frederick~J Stimler.
	\newblock Nitinol characterization study.
	\newblock {\em NASA}, CR-1433, 1969.
	
	\bibitem{philen2007fluidic}
	Michael Philen, Ying Shan, Kon-Well Wang, Charles Bakis, and Christopher Rahn.
	\newblock Fluidic flexible matrix composites for the tailoring of variable
	stiffness adaptive structures.
	\newblock In {\em 48th AIAA/ASME/ASCE/AHS/ASC Structures, Structural Dynamics,
		and Materials Conference}, page 1703, 2007.
	
	\bibitem{jin2010uv}
	Jing Jin, Shuangjun Chen, and Jun Zhang.
	\newblock Uv aging behaviour of ethylene-vinyl acetate copolymers (eva) with
	different vinyl acetate contents.
	\newblock {\em Polymer degradation and stability}, 95(5):725--732, 2010.
	
	\bibitem{boubakri2010impact}
	A~Boubakri, Nader Haddar, K~Elleuch, and Yves Bienvenu.
	\newblock Impact of aging conditions on mechanical properties of thermoplastic
	polyurethane.
	\newblock {\em Materials \& Design}, 31(9):4194--4201, 2010.
	
	\bibitem{lagoudas2008shape}
	Dimitris~C Lagoudas.
	\newblock {\em Shape memory alloys: modeling and engineering applications}.
	\newblock Springer, 2008.
	
	\bibitem{keim2019memory}
	Nathan~C Keim, Joseph~D Paulsen, Zorana Zeravcic, Srikanth Sastry, and Sidney~R
	Nagel.
	\newblock Memory formation in matter.
	\newblock {\em Reviews of Modern Physics}, 91(3):035002, 2019.
	
	\bibitem{paik2012robotic}
	Jamie~K Paik, An~Byoungkwon, Daniela Rus, and Robert~J Wood.
	\newblock Robotic origamis: Self-morphing modular robot.
	\newblock In {\em ICMC}, 2012.
	
\end{thebibliography}

\end{document}


\title{Supplementary Information - Supervised learning in a mechanical system}

\author{Menachem Stern, Chukwunonso Arinze, Leron Perez, Stephanie Palmer, Arvind Murugan}

\affiliation{Physics Department and the James Franck Institute,
	University of Chicago, Chicago, IL 60637}


\pacs{}
\maketitle

\section*{Supplementary Note 1 - Folding origami sheets}

\subsection*{Energy of folded structures}

The origami sheets used in this work are based on a self-folding origami energy model developed and validated in previous studies~\cite{hull:2002,Tachi:2010ge,Pinson2017-td}. The effects of stiff creases are modeled by using torsional spring elements on each crease~\cite{stern2017complexity,stern2018shaping}. Here we discuss in detail how the energy of a folded structure is computed.

For thin origami sheets with free-folding creases, the primary contribution to the energy of a folded structure is due to bending of the sheet faces. Instead of modelling the faces directly, we look at the mechanical constraints inherent to the geometry of the vertices. An origami vertex is known to apply $3$ constraints on the dihedral folding angles of the creases connected to it (due to to embedding of the sheet in $3d$-space). The constraints can be derived by noting that the vertex must not tear open when folded. Thus, starting from any crease, alternating rotations about the dihedral and sector angles around the vertex have to result in an identity operation~\cite{Tachi:2010ge, stern2017complexity, stern2018shaping}.

Suppose there are $N$ creases denoted by an index $i$, each forded to an angle $\rho_i$, and $N$ sectors with angles $\theta_i$ around the vertex. Rotations about one dihedral angle and one sector would combine to form a rotation matrix

\begin{equation} R_i = 
\begin{pmatrix} 1 & 0 & 0 \\ 0 & \cos \rho_i & -\sin \rho_i \\ 0 & \sin \rho_i & \cos \rho_i \end{pmatrix}
\begin{pmatrix} \cos \theta_i & -\sin \theta_i & 0 \\  \sin \theta_i & \cos \theta_i & 0\\ 0 & 0 & 1 \end{pmatrix}.
\label{eqn:matrix}
\end{equation}

For the vertex to be closed (i.e. not torn open) in a folded structure, the combination of rotation about all crease dihedral angles and sector angles must be the identity:

\begin{equation}
A\equiv\prod_{i=1}^{n}R_i = I.
\label{eqn:constraints}
\end{equation}

A folded structure with values $\rho_i$ that do not satisfy Eq.~\ref{eqn:constraints} must cause the sheet faces to bend. Mathematically, this effect will manifest in finite off-diagonal values in the matrix $A\equiv \prod_{i=1}^{N}R_i$. As there are $3$ independent non-diagonal elements, we say that the vertex imparts $3$ mechanical constrains on the dihedral angles $\rho_i$ around it. 

At the flat state all $\rho_i=0$ all constraints are trivially satisfied, so we can write down an expansion for the $3$ off-diagonal terms of $A$ ($T_1\equiv A_{12},T_2\equiv A_{13},T_3\equiv A_{23}$) in powers of the folding angles:

\begin{equation}
T_a(\rho_i) = C_{a}^i \rho_i + D_{a}^{ij} \rho_i \rho_j + \ldots
\label{eqn:expansion3}
\end{equation}

Then, the energy of breaking these constraints is taken as the sum of squares of the residues $T_a$ of the constraint equations $E_{\text{Vertex}} \sim \sum_a T_a(\rho_i)^2$. Summing this vertex energy over all the vertices of the sheet gives rise to the total face bending energy. The energy due to folding of a stiff crease (modeled as a torsional spring with modulus $\kappa_i$) is quadratic in the folding angle $E_{\text{Crease},i}=\frac{1}{2}k_i\rho_i^2$. The total energy of a folded sheet with stiff creases is thus computed as

\begin{equation}
\begin{aligned}
E_{\text{sheet}}(\bm{\rho}) &\equiv E_{\text{Face}} + E_{\text{Crease}} = \sum_{v\in vertices} \sum_{a=1}^3 T_{va}(\bm{\rho}_v)^2 + \frac{1}{2}\sum_{i\in \text{creases}}k_i\rho_i^2.
\end{aligned}
\label{eqn:EnergyModelSI}
\end{equation}

Note that we do not assign an explicit energy scale to the face bending term, so it is implicitly combined with the crease stiffness scale $\bar{k}$.

\subsection*{Folding protocol}

Now that the energy of every folded structure $\rho_i$ of a specific sheet is defined. We can use this energy landscape to simulate the folding of the sheet. Experimentally there are multiple different ways to fold origami sheets~\cite{Peraza:2014}, and we have previously outlined how these methods can be simulated numerically~\cite{stern2018shaping}. 

One way that an origami sheet can be folded is by applying torques directly to the different creases. Suppose a crease $i$ of a flat sheet is subjected to an external torque $F^{ext}_i$. Such a torque will induce folding in the crease, but the sheet generally resists folding due to the extra energy that might be associated with a folded structure. Assuming that the folding process is over-damped, we may write a dynamical folding equation

\begin{equation}
\tau_{\text{relax}}\frac{d \rho_i}{dt} = - \frac{\partial E_{\text{sheet}}(\bm{\rho})}{\partial \rho_i}  + F_i^{\text{ext}}
\label{eq:SIFolding},
\end{equation}

where $\bm{\rho}$ is the current folded structure, and $\tau_{\text{relax}}$ a time scale of the over-damped dynamics. In this work we utilize a specific way of folding the origami sheets. Suppose a set of external torques $\bm{F}^{ext}$ is given (this could be a training or a test example as described in the main text). First, the sheet is folded very fast with a strong external torque $\bm{F}^{ext}$, until a certain folding magnitude $\rho\equiv\vert\vert \bm{\rho}\vert\vert$ is reached. For fast folding we can initially disregard the sheet energy and thus get to a state $$\bm{\rho}_{\text{fast}}=\rho\frac{\bm{F}^{ext}}{\vert\vert\bm{F}^{ext}\vert\vert}.$$ 
Then the sheet is relaxed subject to the constraint that the overall folding magnitude is fixed (i.e. finding an energy minimum on a hyper-sphere of radius $\rho$ in $\rho$-space):

\begin{align}
\begin{aligned}
& \underset{\rho_i}{\text{minimize}}
& & E_{\text{sheet}}(\bm{\rho})\\
& \text{subject to}
& & \vert\vert \bm{\rho}\vert\vert = \rho .
\end{aligned}
\label{eq:SImin}
\end{align}

Finding a local minimum on the hyper-sphere guarantees that this folded structure would naturally occur if the sheet is folded with appropriate torques, as any neighboring configuration costs more energy, and the local minimum will attract the folding process. This algorithm is used to mimic experimental fast folding of origami sheets, followed by clamping of a crease at a specific folded dihedral angle. Here we also adjust the clamped angle such that the overall magnitude of folding $\rho$ remains fixed and different (discrete) folded structures may be compared more easily. Such fast folding was tested extensively~\cite{stern2018shaping}, and found to obtain the same results as numerically solving the ODE of Eq.~\ref{eq:SIFolding}.

\subsection*{Origami sheets and applied force patterns}

In this project we use specific self-folding origami sheets. These are triangulated thin sheets, chosen to have the property of self-foldability. As discussed above, a single vertex induces $3$ mechanical constrains on the angles of creases surrounding it. Thus each vertex has to connect at least $4$ creases or it would be locally rigid. On top of that, for a sheet to self-fold, it needs to have one global degree of freedom, so that the number of creases needs to be one more than the number of constraints. 

\begin{figure*}	
\includegraphics[width=1.0\linewidth]{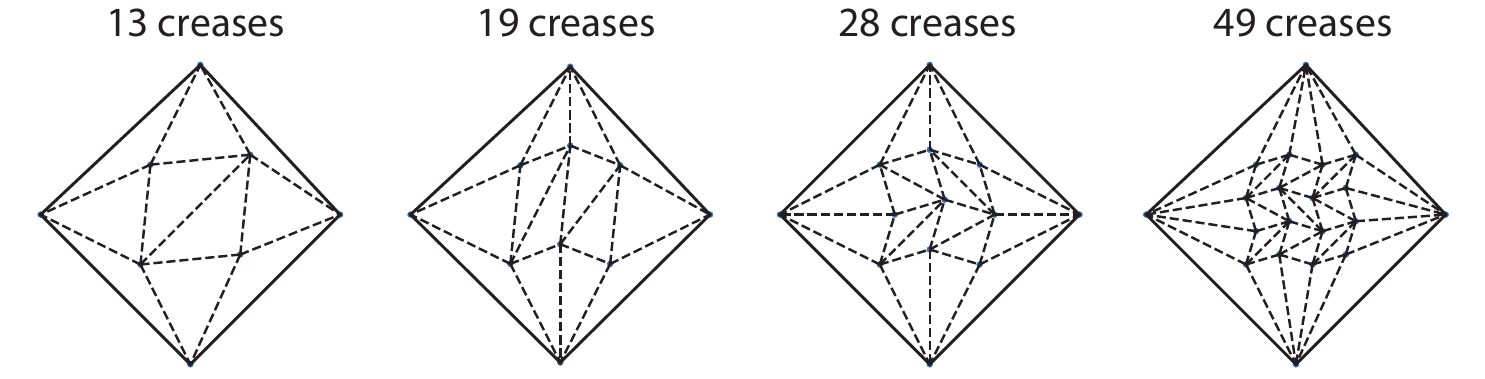}
\caption{Origami Sheets used for training. The size of each sheet is determined by the number creases.
\label{fig:SI-Patterns}}
\end{figure*}

A simple way of generating patterns meeting these requirements is shown in Supplementary Fig.~\ref{fig:SI-Patterns}. These are $4$ specific geometries used throughout this work as the sheets to be trained. Note that we label them according to their size, given by the number creases in each sheet. The number of creases in these sheets are $13,19,28,49$ and the numbers of internal vertices are $4,6,9,16$. Subtracting $3$ times the number of vertices from the number of creases leaves us with one global degree of freedom for each of these sheets, as required.

The number of supported folded structures for these sheets grows exponentially with the number of internal vertices, such that these sheets can fold in approximately $2^4,2^6,2^9,2^{16}$ distinct ways~\cite{stern2017complexity,chen2018branches}. In fact, any sheet with these topologies (yet different geometries) will have a similar number of distinct folded structures. The exact details of the supported folded structures is dependent on the specific geometry, but we only require the existence of many distinct folded structures for the purpose of training. 

These specific sheets, used for training classifiers throughout this work, are definitely not special. We attempted training classifiers using sheets with different geometries and obtained comparable results. In analogy to learning algorithms, the details of the sheet and its supported folded structures correspond to the family of models that the training protocol selects from. For origami, we believe the available classification models are given by merger of attractors of folded structures, supported by the sheet. Since the number of available models to choose from is exponentially large, we reason that the geometry of the sheet should play little role in the success of classification. Therefore, any self-folding origami sheet could be used for training classifiers.

The choice of force patterns applied to the sheets is constrained by the problem definition as training and tests sets. Still, there is usually freedom in how these forces are applied. For example, suppose we wish to train the 13 crease sheet of Supplementary Fig.~\ref{fig:SI-Patterns} on $2d$ force distributions, such as the spherical caps shown in Fig.~3 in the main text. The training and test sets could thus be supplied as pairs of numbers, together with a label (blue\textbackslash orange).
A simple choice for training on such a data set is to pick two creases in the sheet and apply torques directly to these creases, as in Eq.~(\ref{eq:SIFolding}). Here we utilize a different approach.

For an untrained sheet with homogeneous stiffness, it is known that all folded structures reside in the linear null space of the vertex constraint matrix $C$ at the flat state~\cite{stern2017complexity}. Thus, forces applied in a direction within this null space are more `natural' for the sheet, and in general cost much less energy due to face bending. We compute the span of the null space for each one of these sheets, and find that the dimension of the null space is $d_{NS}=\#_{\text{creases}}-2\#_{\text{vertices}}$. Therefore the 13 crease sheet has a $5d$ null space, while the 49 crease sheets has $17d$ null space. Then, the training and test examples are mapped to forces in the null space as follows. For a $n-d$ data set, we choose $n$ random orthonormal vectors in the null space. Each training\textbackslash test example is mapped to a force pattern by assigning every component to one of the random orthonoraml vectors. Now these forces can be directly applied to the sheet to facilitate the training protocol.

Training results in heterogeneous crease stiffness values that change the geometry of the folded structures, so that they do not strictly reside in the null space of the untrained sheet. Still, for the moderate heterogeneity developed during training, the observed folded structures are very close to the null space, such that the described mapping is still useful and practical. 

\section*{Supplementary Note 2 - Training origami sheets}

\subsection*{Learning rule}

As discussed in the main text, self-folding origami sheets naturally give rise to complex mapping of force patterns to folded structures, with exponentially many structures supported by the sheet. The learning rule developed in this work is meant to modify that map by changing crease stiffness coefficients, such that only a small number of folded structures are retained, corresponding to the desired classes. Here we will define precisely how the learning rule is chosen and applied to the sheet in order to develop the desired mapping.

According to the specification of the classification problem, the trainer has no a-priori knowledge of the true underlying force distributions. Instead they are supplied with a list of labeled force patterns (`cats' and `dogs'). These training examples are used to find a reference folded structure in the following way. We fold an untrained sheet with every `dog' example in the training set and record the folding angles of the obtained folded structures. Then, a reference `dog' structure $\hat{\bm{\rho}}_{\textrm{dog}}$ is defined as the average of all of these folded structures (normalized appropriately)

\begin{align}
\begin{aligned}
& \hat{\bm{\rho}}_{\textrm{dog}} \equiv \frac{\sum_{\bm{F}\in \mathcal{F}^{\textrm{dog}}} \bm{\rho}_U(\bm{F})}{\vert\vert \sum_{\bm{F}\in \mathcal{F}^{\textrm{dog}}} \bm{\rho}_U(\bm{F})\vert\vert} ,
\end{aligned}
\label{eq:RefDog}
\end{align}

with  $\mathcal{F}^{\textrm{dog}}$ the set of `dog' training force patterns and $\bm{\rho}_U(\bm{F})$ the folded response of the untrained sheet to force pattern $\bm{F}$. A similar reference state $\hat{\bm{\rho}}_{\textrm{cat}}$ is obtained for the `cat' training examples. Crucially, once the reference structures are set for the untrained sheet, they are kept fixed throughout the training process. These reference structures are used to define the learning rule discussed in the main text. Suppose that during the training protocol, we choose a random `dog' example $\bm{F}^\textrm{dog}$ and apply it to the sheet. The normalized resulting folded structure is written as $\bm{\rho}(\bm{F}^\textrm{dog})$. The learning rule then compares this folded structure to the reference structures defined above and the stiffness coefficients are modified as follows:

\begin{align}
\begin{aligned}
& \text{if } \bm{\rho}(\bm{F}^\textrm{dog}) \cdot \hat{\bm{\rho}}_{\textrm{dog}} > \bm{\rho}(\bm{F}^\textrm{dog}) \cdot \hat{\bm{\rho}}_{\textrm{cat}} \text{ : }
& & \frac{dk_i}{dt} = - \alpha \rho_i^2(\bm{F}^\textrm{dog})\\
& \text{else }  \text{ : }
& & \frac{dk_i}{dt} = + \alpha \rho_i^2(\bm{F}^\textrm{dog})\\
& k_i \geq 0, \; i \in \text{creases}\
\end{aligned}.
\label{eq:SI-LearningRule}
\end{align}

In essence, the learning rule checks whether the observed folded structure is closer to the `dog' reference then to the `cat' reference. If it does, the stiffness of creases that fold considerably in that structure is reduced, effectively reinforcing this force-fold mapping. An opposite modification occurs if the folded structure is far away from the `dog' reference. A similar training rule is used when `cat' forces patterns are applied, with the understanding that we wish to compare the resulting folded structure $\bm{\rho}(\bm{F}^\textrm{cat})$ to the `cat' reference $\hat{\bm{\rho}}_{\textrm{cat}}$.

\subsection*{Assigning labels to folded structures}

To begin with, we are given labeled force patterns, and an untrained sheet with many available folded structures. It is important to note that these folded structures are equivalent and not intrinsically labeled. Thus, as part of the learning protocol we must specify how to label these folded structures, and in particular which of them to call `dog' and `cat' (or `blue' and `orange'). A simple solution would be to choose $2$ of the folded structures in advance and assign the classification labels to them. Unfortunately, this turns out to be too restrictive for a couple of reasons. First, the choice may be far from ideal in the sense that these labeled folded structures are very different than the actual folded response of the sheet to the labeled force patterns. Furthermore, as the training process modifies the stiffness of different creases, the folded structures supported by the sheet change as well, either by moving around or disappearing altogether in saddle-node bifurcations~\cite{stern2018shaping}. We thus take a different approach to labeling folded structures, as detailed below.

Suppose we have trained a sheet for some time, and it now has a particular stiffness profile on its creases $k_i$. To find a folded structure of this sheet to be labeled `dog', we apply each of the `dog' training examples once, and record the discrete resulting folded structures due to all of them $\{\bm{\rho}(\bm{F}\in\mathcal{F}^{\text{dog}})\}$. We then count the training force patterns that folded into each one of the structures in this set. The folded structure that resulted from the largest number of training force patterns is chosen to be labeled as `dog'. In case of a tie, e.g. two or more folded structures folding as a result of the same number of force patterns, one of these structures is randomly chosen to serve as the label. Thus, the labels for `dog' and `cat' are decided through simple plurality rules every time we compute the classification accuracy of the sheet. Note that force patterns may also fold the sheet into structures not labeled as either `cat' or `dog', in which case they count as failed classification. If both `dog' and `cat' labels are chosen to be associated with the same folded structure, a plurality rule between the two classes decides which class is labeled with that structure (i.e. whether more `cat' or `dog' force patterns folded into that structure), while the other is assigned with the runner up structure of that labels' plurality vote. Finally, if the sheet is over-trained to the point where only one folded structure remains, that structure is labeled as both `cat' and `dog', such that classification fails completely, by definition.

\subsection*{Effective cost function}

In this work we have defined our learning rule as a supervised physical process modifying the stiffness coefficients of an origami sheet. It is interesting to compare this kind of learning protocol to more established learning algorithms originating in computer science and statistics. One important difference is that traditional learning algorithms are usually defined as an optimization problem, where the function to be optimized (often called cost or loss function) incorporates the training data. 

A simple example is linear regression, where the cost function is usually chosen as a least squares form, where the differences are taken between a linear model $h(x)$ and the observations $y$:

\begin{align}
\begin{aligned}
\text{Cost} &\equiv \sum_{d\in \text{data}} (h(x_d)-y_d)^2 \\
h(x) &= a_0 + a_1 x
\end{aligned}.
\label{eq:SI-LinReg}
\end{align}

The regression (or learning algorithm) then optimizes the cost function with respect to the model parameters $\bm{a}\equiv (a_0,a_1)$  
\begin{align}
\begin{aligned}
& \underset{\bm{a}}{\text{minimize}} 
& & \text{Cost}(\{x\},\{y\};\bm{a}) \nonumber .
\end{aligned}
\end{align}

This optimization can be achieved in any number of ways, but a practically favored method (at least for more advanced algorithms like deep learning) is mini-batch stochastic gradient descent (SGD)~\cite{ruder2016overview}. In an extreme case, when the mini-batches are chosen to be of size $1$, a single training example $(x,y)$ is chosen at random in each step, and one computes the gradient (with respect to parameters $\bm{a}$) of the cost function defined with this example alone $\bm{G}\equiv \nabla_{\bm{a}} (h(x)-y)^2$. Now, training proceeds by modifying the parameters in proportion to the the gradient of this single example cost function

\begin{align}
\begin{aligned}
\bm{a} \rightarrow \bm{a} - \alpha \bm{G}
\end{aligned},
\label{eq:SI-Descent}
\end{align}

where $\alpha$ is a scalar known as the learning rate. We may compare this single example SGD with our origami training protocol. It is relatively easy to see that our training rule (Eq.~\ref{eq:SI-LearningRule}), once a standard wait time is chosen at the folded state, has the form of SGD, making it similar in essence to other learning algorithms. To find out what effective cost function gives rise to the origami learning rule, we integrate Eq.~\ref{eq:SI-LearningRule} with respect to the stiffness coefficients

\begin{align}
\begin{aligned}
&\text{cost}_\text{map}(\bm{\rho}(\bm{F}^\textrm{dog})) = f \sum_{i\in \text{creases}} k_i \rho_i^2(\bm{F}^\textrm{dog}) \\
& \text{if } \bm{\rho}(\bm{F}^\textrm{dog}) \cdot \hat{\bm{\rho}}_{\textrm{dog}} > \bm{\rho}(\bm{F}^\textrm{dog}) \cdot  \hat{\bm{\rho}}_{\textrm{cat}} \text{ : } & & f = +1 \\
& \text{else }  \text{ : }
& & f = -1
\end{aligned}.
\label{eq:SI-OrigamiCost}
\end{align}

Similarly to the linear regression example, our origami training protocol attempts to minimize this derived cost function, one training example at a time. Inspecting this function, note that it is very similar to the energy of the torsional springs in the folded structure $E_{\text{Crease}}(\bm{\rho})\sim \sum_i k_i \rho_i^2$. The difference is in the `supervising factor' $f$ that can be $\pm 1$ whether the folded structure is accepted or not. We conclude that our origami training protocol is attempting to minimize the energy of accepted folded structures, while maximizing the energy of rejected structures.

\section*{Supplementary Note 3 - Using origami sheets to define classification problems}

The force distributions classified in the main text are relatively simple. Both the spherical cap and the Iris data distributions can be well separated by a hyper-plane, a very simple decision boundary. It is interesting to study the type of decision boundaries naturally trainable in origami sheets -- and whether they can be used to classify intrinsically high dimensional data. 

\begin{figure*}	
\includegraphics[width=1.0\linewidth]{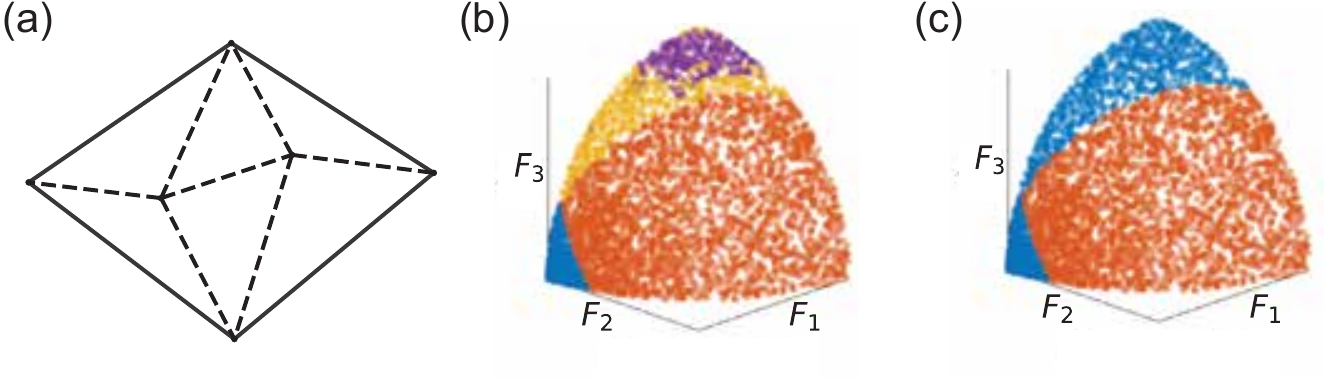}
\caption{Defining force distributions using the force-fold mapping of an origami sheet. a) Origami sheets with $2$ internal vertices support $4$ discrete folded structures. b) Sample force patterns on a $2$-sphere show the force-fold mapping ($4$ color coded regions). c) When some attractor regions are merged (here, blue, yellow and purple are merged), we obtain an intrinsically $2$-dimensional separator surface between two classes of force patterns.
\label{fig:SI-MakeDistrib}}
\end{figure*}

There are many ways to obtain high dimensional distributions. Here we choose to study distributions derived from the folding maps of origami sheets. Consider a relatively simple sheet with $2$ internal vertices (Supplementary Fig.~\ref{fig:SI-MakeDistrib}a). It is known that such sheets support $4$ discrete folded structures, and that the linearized null space in which they reside is $3$-dimensional. Therefore, if we sample random force patterns within this null space, we expect to see the sheet folding into $4$ distinct structures (color coded regions in Supplementary Fig.~\ref{fig:SI-MakeDistrib}b). The forces $F_1,F_2,F_3$ are assigned by randomly choosing Euler angles on the $2$-sphere, and $3000$ data points are sampled on the positive octant.  Note that we sample normalized forces on the surface of a $2-sphere$, such that the distribution of force patterns is actually $2$-dimensional.

Now, suppose we wish to classify forces to $2$ classes (`blue'\textbackslash `orange'). A simple way to create $2$ neighboring sets of points is to take the data of Supplementary Fig.~\ref{fig:SI-MakeDistrib}b and merge some attractor regions to create larger groups of points. In Supplementary Fig.~\ref{fig:SI-MakeDistrib}c, we merge the `blue', `yellow', and `purple' folded structures to create one region we define as `blue'. This process yields two distributions that are intrinsically $2$-dimensional, and not naturally separable by a hyper-plane. Larger sheets can be similarly used to create force distributions in higher dimensional space.

\begin{figure*}	
\includegraphics[width=1.0\linewidth]{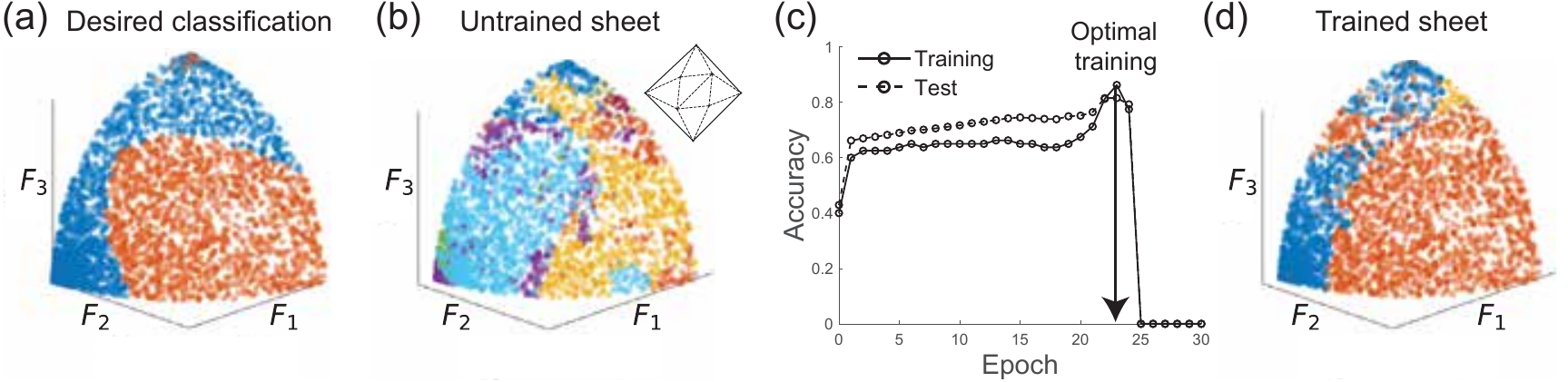}
\caption{Training a sheet on a force distribution derived from a different sheet. a) Target classification, a sample distribution derived from a small, $2$-vertex origami sheet. b) The force-fold map of an untrained 13 crease sheet is very different from the desired mapping. c) With training, the accuracy of classification improves and peaks at $82\%$. d) The optimally trained sheet has a complex decision boundary that resembles (but different than) the desired boundary.
\label{fig:SI-HighD}}
\end{figure*}

With this process, we have access to a new variety of $2$-way classification problems, on which we can try to train origami sheets using the training protocol described in the main text. Crucially, the sheet used to classify such distributions is different than the sheet used to derive the distribution. In other words, we ask if our training protocol can induce an origami sheet to mimic the force-fold mapping of another sheet.

Suppose we want to classify the distribution seen in Supplementary Fig.~\ref{fig:SI-HighD}a, derived form a $2$-vertex sheet as described above. We wish to train a 13 crease sheet to classify this force pattern data. The untrained sheet has $2^4$ discrete folded structures that do not align with the target distribution in any representation that we tested (Supplementary Fig.~\ref{fig:SI-HighD}b). The problem of classification here is to train this sheet to have just $2$ folded structures with the right force-fold mapping as in the target distribution. 

The target distribution is mapped to applied force patterns on the 13 crease sheet by the construction describe in Supplementary Note 1: choosing random orthonormal vectors in the null space of the 13 crease sheet and mapping the distribution as components of these vectors. We then randomly sample $20$ `blue' and $20$ `orange' force patterns, marked as diamonds in Supplementary Fig.~\ref{fig:SI-HighD}, to serve as the training set. As we train the sheet, the classification accuracy improves dramatically and reaches a maximum of $82\%$ (test accuracy) after $23$ epochs (Supplementary Fig.~\ref{fig:SI-HighD}c). To qualify the classification better, we look at the classification results corresponding to the maximal accuracy at epoch $23$ (Supplementary Fig.~\ref{fig:SI-HighD}d).  We observe that the trained decision boundary resembles the desired boundary, so that the training protocol indeed produced a reasonable classification. 

Note a few artifacts that still remain in the trained map: 1) there are 3 folded structures left, rather than 2 (a small third color coded region exists, labeled yellow), 2) a second orange region appeared inside the bulk blue region, emphasizing that the decision boundaries between folded structures in sheets are generally \textit{not} hyper-planes. We conclude that origami sheets can be trained to classify distributions derived from other sheets, that are intrinsically higher dimensional than the problems discussed in the main text. We leave questions of the sheet size and the complexity of decision boundaries to future studies.

\section*{Supplementary Note 4 - Transforming Iris data to applied forces on sheets}

The Iris data set~\cite{Iris} classified in the main text is a classical problem for classification. In this work we are able train an origami sheet to correctly classify two species of Iris (I. Versicolor, I. Virginica) at an accuracy of $91\%$. Here we discuss how the Iris data is used to generate training and test sets of applied force patterns to be used on origami sheets.

Each Iris example in the data set is given as a vector with $4$ features (components): sepal length, sepal width, petal length, petal width. These length measurements are all given in $cm$. In addition to these measurements, each Iris is labeled as one of the Iris species in the study. To generate force pattern sets from this data, we would like the different measurements for each Iris example to be components of force vectors in the null space of the origami sheet, as described in Supplementary Note 1. However, the raw Iris data is not suited for this purpose due to two reasons. The dimensionful measurements of Iris lengths, if directly translated to forces, would be far too great for our sheets and will cause it to fold too much and cause the sheet faces to collide. More crucially, sepal and petal lengths tend to be considerably larger than their widths, and the same goes for the variance of these variables. This will causes the width variables to be perceived as less important in the training protocol, and have a negative effect on the classification results.

Fortunately, diverse data like this is an issue regularly faced by learning algorithms, and it is generically solved by applying an invertible transformation to the data. The transformed data is then better suited for the learning algorithm in use. A typical example of such a transformation in data sets is to normalize each feature (divide by the mean of that feature) and translate it such that the mean of the transformed data is $0$. This transformation is especially useful for classification algorithms like logistic regression, where the different features have different dimensional units. 

In our case however, the standard transformation above is not useful, due to a particular property of origami sheets, namely their $Z_2$ symmetry. If forces $\bm{F}$ are applied to the sheet and it folds into a state $\bm{\rho}$, then folding the same sheet with forces $-\bm{F}$ will result in a state $-\bm{\rho}$. This is true for any self-folding origami sheet, regardless of the stiffness profile on its creases. This property cannot be changed by training the sheet. Thus, force patterns of opposite sign and different labels cannot be correctly classified. A simple way to avoid this issue is to limit the force patterns to reside in a restricted part of force space. We choose to limit the distributions such that the transformed Iris data will all be in the positive $4$-hyperoctant. 

In addition, we want the data to span as much as possible of the positive hyperoctant. This will increase the expressive of our training protocol, as more discrete folded structures would become available if the applied force patterns are more diverse. We thus need to transform the Iris data to be all positive, and stretch it such that all features have similar variance.

To achieve these goals we apply the following linear (invertible) transformation to the Iris data of the Versicolor and Virginica species. Suppose an Iris example is given as a vector $\bm{x}$ (where the components are sepal length, sepal width, petal length, petal width in this order). The vector is transformed by

\begin{align}
\bm{x}^* = A\bm{x} + b& \nonumber \\ 
A = \begin{pmatrix} 0.264 & 0 & 0 & 0 \\ 0 & 0.580 & 0 & 0 \\ 0 & 0 & 0.303 & 0 \\ 0 & 0 & 0 & 0.836 \end{pmatrix}&\ \  , \ \ b = -0.880 . 
\label{eqn:IrisTrans}
\end{align}

Then the transformed vector is used to define the force patterns applied to the origami sheet, as described in Supplementary Note 1. After training is concluded, the transformation can be inverted to relate the origami classification results with the original Iris data, as shown in the main text.
